\documentclass[11pt,a4paper]{book}

\newcommand{\bi}[1]{\mbox{\boldmath$#1$}}
\newcommand{\balpha}{\mbox{\boldmath$\alpha$}}
\newcommand{\bbeta}{\mbox{\boldmath$\beta$}} 
\newcommand{\bsigma}{\mbox{\boldmath$\sigma$}} 
\newcommand{\bmu}{\mbox{\boldmath$\mu$}} 
\newcommand{\bomega}{\mbox{\boldmath$\omega$}} 

\begin{document}

\chapter*{Generalized Lagrangians and spinning particles}

\begin{center}
{\bf Martin Rivas}\\
Theoretical Physics Department\\
The University of the Basque Country\\
Apdo. 644, Bilbao, Spain.\\
E-mail:wtpripem@lg.ehu.es
\end{center}

\vspace{1cm}

\begin{quotation}\noindent{\bf Abstract.-}
The use of generalized Lagrangians for describing elementary particles
was already claimed by Ostrogradskii. It is shown how the spin structure
of elementary particles arises if one allows the Lagrangian to depend on higher
order derivatives. One part is related to the rotation of the particle and the other,
which is coming from the dependence of the Lagrangian on the acceleration, is known as the
zitterbewegung part of spin.
\end{quotation}

\section*{Introduction}

As early as 1736 Leonard Euler presented the book {\sl Mechanica} in which he
established for the first time Newtonian dynamics in terms
of rational mechanics. At that time Euler's law of mechanics took the form: 
\begin{quote}{\sl the increase $dv$
of the velocity is proportional to $pdt$, where $p$ is the power acting on the body
during the time $dt$.}\footnote{R. Dugas, {\sl A History of Mechanics}, Dover NY (1988).}
\end{quote}
Later, in 1740, in his work {\sl Methodus inveniendi lineas curvas},
introduced the calculus of variations, fundamental in the subsequent development of analytical dynamics.
He stated the variational principle in the form: 
\begin{quote}
{\sl Since all the effects of Nature obey some 
law of maximum or minimum, it cannot be denied that the curves described by projectiles 
under the influence of some forces will enjoy the same property of maximum or minimum. 
It seems easy to define, a priori, using metaphysical principles, what this property is. But since,
with the necessary application, it is possible to determine these curves by the direct
method, it may be decided which is a maximum or a minimum.}
\end{quote}
The magnitude which he considered to be stationary was $mds\sqrt{h}$, where $m$ is the mass of the body,
$ds$ the element of distance travelled and $h$ the height of fall.
It was in 1749 at the Academy of Sciences of Berlin when he presented Newton's law
in the standard form ${\bi f}=m{\bi a}$. Explicitly
\[
I.\;\frac{2ddx}{dt^2}=\frac{X}{M},\quad
II.\;\frac{2ddy}{dt^2}=\frac{Y}{M},\quad
III.\;\frac{2ddz}{dt^2}=\frac{Z}{M},
\]
where $X$, $Y$ and $Z$ are the Cartessian components of the external 
force, and the left hand sides are a peculiar form of writting the second derivative
of the position variables.
Euler left Berlin and moved to St. Petersburg in 1766 where he wrote as much as
half of his extensive work and where he died in 1783.

Joseph Louis Lagrange succeeded Euler in 1766 as director of mathematics in the Academy of Sciences of Berlin. 
In 1787 became a member of the Paris Academy of Sciences where he published in 1788 the book
{\sl M\'ecanique Analytique} in which the methods of Lagrangian dynamics
were introduced. He said:
\begin{quote}
{\sl In the general motion of any system of bodies, actuated by mutual forces of attraction, or by attraction
towards fixed centres which are proportional to any function of the distance, 
the curves described by the different bodies, and their velocities, are necessarily 
such that the sum of the products of each mass by the integral of the product of the velocity
and the element of the curve is necessarily a maximum or a minimum; provided that the first 
and last points of each curve are regarded as fixed, so that the velocities 
of the corresponding coordinates at those points are zero.\footnote{J.L. Lagrange, 
{\sl M\'ecanique Analytique}, vol I, p. 276.}}
\end{quote}
In contrast with Newton's {\sl Principia} in which many geometrical diagrams are used to produce the corresponding
proofs, Lagrange enhanced the role of analysis and made the following declaration
\begin{quote}
{\sl No diagrams will be found in this work. The methods that I explain in it require neither constructions
nor geometrical or mechanical arguments, but only the algebraic operations inherent to a regular and uniform process.
Those who love Analysis will, with joy, see mechanics become a new branch of it and will be grateful to me for
thus having extended this field.
}
\end{quote}
As a consequence of Lagrange's equations Mechanics rested on the Principle of 
least action, or as stated above
as the {\sl principle of the greatest or least living force}. 

Mikhail Vasilevich Ostrogradskii, whose bicentennial we are celebrating, left Ukraine at the age of 21 
to study in Paris. 
Between 1822 and 1827 he attended lectures by Laplace, Fourier, 
Legendre, Poisson and Cauchy and publish several papers in the Paris Academy. He
went to St Petersburg in 1828. Since then, Ostrogradskii lectured at the Naval Academy, 
from 1830 at the Institute of Communication and, from 1832, also at the Pedagogical Institute. 
In 1847 he became chief inspector for the teaching of mathematical sciences in military schools. 
He established the conditions which allowed Chebyshev's school to 
flourish in St Petersburg. 
He should also be considered as the founder of the Russian school of theoretical mechanics. 
I am not an expert in the History of Physics, but needless to say that Ostrogradskii was probably
impregnated of Euler and Lagrange ideas during his stay in Paris and St Petersburg.
He was aware of the importance of variational methods in mechanics and the Lagrangian 
formalism in particular.
We are concerned in this contribution with his suggestion in 1850 of using Lagrangians depending on 
higher order derivatives~\footnote{M. Ostrogradskii, {\sl Mem. Acad. St. 
Petersburg}, {\bf 6}(4), 385 (1850).} and their usefulness to describe classical spinning particles.

\section*{What is classical spin?}

Because Newton's equations for point particles are second order ordinary 
differential equations for the position variables of the particles,
the action principle can be written in terms of the Lagrangian function, which is therefore, an explicit
function of the position variables and their first order time-derivatives. In rational mechanics
this has been extended to arbitrary systems in the sense that Lagrangians are postulated as 
functions of the independent degrees of freedom $q_i$ and also of only their first order derivatives. 

But this, which is world-wide
accepted as the basis of point particle dynamics, will no longer be considered as such for the description
of elementary spinning particles. In general, classical spin is considered as some kind of vector of constant 
magnitude, attached to the point particle, constant in time for the free motion and when some interaction
is present it experiences some precession due to the torques of the external forces. But this idea
of a classical spin is difficult to agree with the notion of electron's spin when 
considered under the analysis of Dirac's equation.\footnote{P.A.M. Dirac, {\sl The Quantum Theory of the 
Electron}, Proc. Roy. Soc. Lon. {\bf A117}, 610 (1928); {\bf A118}, 351 (1928); {\sl The Principles of Quantum 
mechanics}, Oxford Univ. Press, 4th ed. Oxford (1967).} 
According to Dirac, the spin for the free electron satisfies the dynamical equation
\begin{equation}
\frac{d{\bi S}}{dt}={\bi P}\times{\bi u},
\label{eq:s}
\end{equation}
where 
\[
{\bi S}=\frac{\hbar}{2}\pmatrix{\bsigma&0\cr 0&\bsigma},
\]
is the spin operator expressed in terms of Pauli's $\bsigma$ matrices, ${\bi P}$ is the total 
constant linear momentum of the electron and ${\bi u}=c\balpha$ is Dirac's velocity operator
written in terms of Dirac's $\balpha$ matrices. If we take the scalar product with ${\bi S}$ of equation (\ref{eq:s})
we get
\[
{\bi S}\cdot\frac{d{\bi S}}{dt}=\frac{1}{2}\,\frac{d S^2}{dt}={\bi S}\cdot({\bi P}\times{\bi u})\neq0,
\]
and therefore neither the spin nor its absolute value are constants of the motion for a free electron.

It is clear that the idea of a classical spin as a kind of pin stuck to the point particle 
which remains constant whenever the particle is free and 
satisfies some plausible dynamical equation when interacting, must be abandoned or at least revisited.
Therefore, if we do not know what are the necessary 
variables to describe spin at the classical level we do not know whether they must satisfy or not 
second order differential equations. 

This has been one of the leading arguments to a thorough revision of kinematics and dynamics,
to consider the possibility of a classical description of spin. This has been done 
in previous works\footnote{M. Rivas, {\sl Classical Particle Systems: I. 
Galilei free particles}, J. Phys. {\bf A 18}, 1971 (1985);
{\sl Classical Relativistic Spinning Particles}, 
J. Math. Phys. {\bf 30}, 318 (1989).},
and a more detailed and comprehensive analysis is collected in the 
book \footnote{M. Rivas, {\sl Kinematical theory of spinning particles},
Kluwer, Dordrecht, (2001).}. A second important argument has been the strength given in Lagrangian dynamics 
to the end point variables of the variational formalism, as was already considered by Lagrange in the mentioned
statement of the previous section. We shall call {\bf kinematical variables} to the variables that define
these end-points and {\bf kinematical space} to the manifold they span.

\section*{Classical Elementary Particles}

For the revision of a classical description of matter we can take a look at
the successful way quantum mechanics describes both kinematics and dynamics.
By kinematics we understand the basic statements that define the physical objects we 
go to work with and their analytical description. 
In quantum mechanics, a state of an elementary particle is a vector state of
an irreducible representation of the kinematical group of space-time 
transformations that describes the Relativity Principle.~\footnote{\hspace{0.1cm}E.P. Wigner, {\sl Ann. Math.} 
{\bf 40}, 149 (1939). } This is a group theoretical definition of an elementary particle.
Intrinsic attributes of the particles are then interpreted in terms of group invariants and are therefore
related to the Casimir operators of the kinematical group or, properly speaking, to the Casimir
operators of their projective unitary irreducible representations. 

Quantum dynamics describes the probability
amplitudes for a whole process in terms of the end point kinematical
variables that characterize the initial and final
states of the system. The details concerning the intermediate flight of the particles involved,
are not explicit in the final form of the result. They are all removed in the calculation process, enhancing the role,
as far as the theoretical analysis is concerned, of the initial and final data. This looks similar, 
at least in a formal way, to the variational formalism. But when one quantizes a Lagrangian system by means of
Feynman's path integral approach\footnote{R.P. Feynman, R.B. Leighton and M. Sands, {\sl The 
Feynman Lectures on Physics}, vol.~III, Addison-Wesley, Reading, Mass. (1965).}, this probability
amplitude (there called Feynman's kernel) becomes a function of only the initial and final kinematical variables 
of the classical system.

Therefore, in the classical approach, we define a classical elementary particle
by giving a group theoretical characterization to the kinematical variables of the action integral of the Lagrangian.
We postulate the following
\begin{quote}
{\bf Definition.-} A classical elementary particle is a Lagrangian system
whose kinematical space is a homogeneous space of the kinematical group.
\end{quote}

From the mathematical viewpoint the largest homogeneous space of a Lie group is the group itself
and therefore this definition restricts the maximum number of kinematical variables
to the number of group parameters. Any homogeneous space of a group
inherits not only part of the structure of the group it comes from, but at the same time the physical
dimensions of the corresponding parameters. It is the group and the variables we use to parametrize it
which determine the basic variables that define a classical elementary particle, and also
their physical or geometrical interpretation. 

In this way we do not state that Lagrangians depend only on first order derivatives. This method depends only on
the kind of variables we fix as end-points of the variational process. Once the end-points are fixed,
the variational problem requires for the Lagrangian to depend on the next order of 
derivation of all the kinematical variables. It is this construction which will produce, or not,
the dependence on higher order derivatives, so that the restriction of using first order Lagrangians
is a consequence of the particular election of the kinematical variables.

\section*{Mathematical properties of the Lagrangian}

When we have a generalized Lagrangian $L(t,q_i,\ldots,q_i^{(k)})$, which is an explicit function of time $t$,
of the $n$ degrees of freedom $q_i$, $i=1,\ldots,n$ and their derivatives up to some finite 
order $k$, $q_i^{(k)}\equiv d^kq_i/dt^k$,
the kinematical variables are therefore the time $t$,
the $n$ degrees of freedom $q_i$, $i=1,\ldots,n$ and their derivatives up to order $k-1$. 
These are the variables we have to leave 
fixed at the initial time $t_1$ and final time $t_2$ of the action functional. 
We denote them genericaly by $x_l$, $l=0,\ldots,nk$ 
where we take $x_0=t$, $x_i=q_i$, $x_{n+i}=q_i^{(1)}$, and so on. We clearly see that $L$ depends also 
on the next order derivative of the kinematical variables. 

Now, to produce a coherent relativistic formalism we need first to withdraw the time as 
evolution parameter, and consider that evolution is described in terms of some arbitrary
parameter $\tau$. Therefore, the time derivative $q_i^{(1)}=dq_i/dt$ should be replaced by
$q_i^{(1)}=\dot{q_i}/\dot{t}$, $q_i^{(2)}=\dot{q_i}^{(1)}/\dot{t}$, and so on and where the dot
over a variable means its $\tau-$derivative, in such a way
that the action functional can be rewritten as
\[
\int_{t_1}^{t_2}\,L(t,q_i^{(k)}) dt=\int_{\tau_1}^{\tau_2}\,L(t,q_i,\dot{q_i}/\dot{t},\ldots,\dot{q_i}^{(k-1)}/\dot{t})\,\dot{t}\, d\tau\equiv
\int_{\tau_1}^{\tau_2}\,\widehat{L}(x_l,\dot{x}_l)d\tau.
\]

We see from the above change of variables that the new 
Lagrangian $\widehat{L}(x_l,\dot{x}_l)\equiv L\,\dot{t}$, 
written in terms of the kinematical variables, has the following properties:

\noindent{\bf 1.} It is independent of the evolution parameter $\tau$.

\noindent{\bf 2.} It is a homogeneous function of first degree in the derivatives $\dot{x}_l$ of the kinematical
variables, and according to Euler's theorem on homogeneous functions, it satisfies
\[
\sum_j\left(\frac{\partial \widehat{L}}{\partial\dot{x}_j}\right)\,\dot{x}_j=\widehat{L}.
\]

\noindent{\bf 3.} It therefore admits the general form
\begin{equation}
\widehat{L}=\sum_jF_j(x_l,\dot{x}_l)\,\dot{x}_j,\label{eq:homog} 
\end{equation}
where the functions $F_j=
{\partial\widehat{L}}/{\partial\dot{x}_j}$ are homogeneous functions of zero degree in the variables $\dot{x}_l$.

\noindent{\bf 4.} If $G$ is a Lie group of transformations of the kinematical variables $x$, such that
under the transformation $x'=gx$, and the corresponding induced transformation $\dot{x}'=g\dot{x}$ for any $g\in G$, 
the dynamical equations remain invariant, then the Lagrangian
transforms under $G$ as
\begin{equation}
\widehat{L}(gx,g\dot{x})=\widehat{L}(x,\dot{x})+\frac{d\alpha(g;x)}{d\tau},
\label{eq:alfa}
\end{equation}
and where the function ${d\alpha(g;x)}/{d\tau}$ is not arbitrary. It depends only on the kinematical variables
and of the group parameters and it is analyticaly related to the exponents of the group $G$.
\footnote{J.M. Levy-Leblond, {\sl Group theoretical 
foundations of Classical Mechanics: The Lagrangian gauge problem}, Comm. Math. Phys. {\bf 12}, 64 (1969).}

\noindent{\bf 5.} Noether's theorem. 
The invariance of the action functional under an $r-$parameter Lie group $G$, defines $r$ constants of the motion
$N_\alpha$, $\alpha=1,\ldots,r$ which can be written in terms of only the functions $F_l(x,\dot{x})$ and their
first time derivatives and of the first order functions of the infinitesimal transformations of the $x_l$.

\section*{Simple examples}

Let us start first with the Newtonian point particle. By definition 
its kinematical variables for its Lagrangian formalism are time $t$
and position ${\bi r}$. 
Let us assume first that the set of inertial 
observers are all at rest with their
Cartesian frames parallel with respect to each other so that the kinematical group is just
the space-time translation group. Then the kinematical relation 
between observers is given by the group action
 \begin{equation} 
t'(\tau)=t(\tau)+b,\quad {\bi r}'(\tau)={\bi r}(\tau)+{\bi a},
\label{eq:translate}
 \end{equation} 
at any instant of the evolution parameter $\tau$.
This four-parameter group has four 
generators $H$ and ${\bi P}$.
In the action (\ref{eq:translate}) the generators are 
the differential operators 
\begin{equation}
H=\partial/\partial t,\quad {\bi P}=\nabla,
\label{eq:HPdif}
\end{equation}
The group law $g''=g' g$, is
\begin{equation} 
b''=b'+b,\quad {\bi a}''={\bi a}'+{\bi a}.
\label{eq:tras}
 \end{equation} 

We see that the kinematical space of our point particle is in fact 
isomorphic to the 
whole kinematical group, so that our kinematical 
variables $x\equiv(t,{\bi r})$ have the same domains and dimensions as 
the group parameters $(b,{\bi a})$, respectively. The kinematical space $X$ is 
clearly the largest homogeneous space of the kinematical group. 

According to this restricted Relativity Principle, the Lagrangian for a point particle will be a function of the variables $(t,{\bi 
r},\dot t,\dot{\bi r})$, and a homogeneous function of first degree in 
terms of the derivatives $(\dot t,\dot{\bi r})$.  Then, according to (\ref{eq:homog}), 
it can be written as
 \begin{equation}
L=T\dot t+{\bi R}\cdot\dot{\bi r},
\label{eq:Lagpoint}
 \end{equation}
with $T=\partial L/\partial\dot t$ and ${\bi R}=\partial L/\partial\dot{\bi r}$.
Since the space-time translation group has no central extensions and thus no non-trivial
exponents, Lagrangians can be taken strictly invariant under this group. Terefore, 
dynamical equations can be 
any autonomous second order differential equation of the functions 
${\bi r}(t)$, not depending explicitly on the variables ${\bi r}$ and $t$. 

When applying Noether's theorem for this kinematical group we obtain 
as constants of the motion, the energy $H=-T$ and the linear momentum 
${\bi P}={\bi R}$. Possible Lagrangians for this kind of systems are 
very general and might be any arbitrary function of the components of the velocity $dr_i/dt$. 
The homogeneity condition in 
terms of kinematical variables implies that,
for instance, expressions of the form
 \[ 
a_{ij}\frac{\dot r_i\dot r_j}{\dot{t}}+b_{ijk}\frac{\dot{r_i}\dot{r_j}\dot{r_k}}{\dot{t}^2}+\cdots,
 \] 
with arbitrary constants $a_{ij}$, $b_{ijk}$, etc., or expressions like 
\[ 
\sqrt{a_0{\dot{t}}^2+a_{i}\dot{t}\dot{r_i}+b_{ij}{\dot{r_i}\dot{r_j}}+c_{ijk}{\dot{r_i}\dot{r_j}\dot{r_k}}/{\dot{t}}+\cdots\quad}, 
\] 
homogeneous of first degree in the derivatives, can be taken 
as possible Lagrangians. 

Let us go further and extend the kinematical group to include rotations. 
Then, the kinematical transformations are
 \begin{equation} 
t'(\tau)=t(\tau)+b,\quad {\bi r}'(\tau)=R(\bbeta){\bi r}(\tau)+{\bi a},
 \label{eq:Aristotle} \index{kinematical group!Aristotle group}
 \end{equation} 
where $R(\bbeta)$ represents a rotation matrix written in terms of 
three parameters $\beta_i$ of a suitable parametrization of the rotation 
group. This group is called the Aristotle group $G_A$. In addition to $H$ and ${\bi P}$, it has 
three new generators ${\bi J}$, that in the above
action (\ref{eq:Aristotle}) are given by the operators ${\bi J}={\bi r}\times\nabla$.
This group also does not have central extensions, and thus 
no nontrivial exponents. Lagrangians in this case will be also invariant. 
This additional rotation invariance leads to the conclusion that $L$,
which still has the general form (\ref{eq:Lagpoint}),
will be an arbitrary function of $\dot{\bi r}^2$. 

When applying Noether's theorem, we have in addition to the 
energy $H=-\partial L/\partial\dot t=-T$ and linear momentum ${\bi 
P}=\partial L/\partial\dot{\bi r}={\bi R}$, a new observable, related to the invariance under rotations, the 
angular momentum ${\bi J}={\bi r}\times{\bi P}$. 

The group elements are parameterized in terms of the seven parameters 
$g\equiv(b,{\bi a},\bbeta)$ and the group $G_A$ has the composition law $g''=g'g$ 
given by:
\begin{equation} 
b''=b'+b,\quad {\bi a}''=R(\bbeta'){\bi a}+{\bi a}',\quad 
R(\bbeta'')=R(\bbeta')R(\bbeta).
\label{eq:arisgroup}
 \end{equation} 
We clearly see, by comparing (\ref{eq:arisgroup}) with (\ref{eq:Aristotle}), 
that the kinematical space $X$ of our point particle is 
isomorphic to the homogeneous space of the group, $X\simeq G_A/SO(3)$. 
It corresponds to the coset space of elements of the form $(t,{\bi r},{\bf 0})$
when acting on the subgroup $SO(3)$ of elements $(0,{\bf 0},\bbeta)$. The kinematical variables
$(t,{\bi r})$ span the same manifold and have the same dimensions as the set of group elements 
of the form $(b,{\bi a},{\bf 0})$.

But once we have a larger symmetry group, we can extend our definition of 
elementary particle to the whole group $G_A$. The physical system 
might have three new kinematical variables $\balpha$, the angular variables.
In a $\tau$-evolution description of the dynamics, 
with the identification in $g''=g'g$ of 
$g''\equiv x'(\tau)$, $g\equiv x(\tau)$ and $g'$ playing the role of the group element $g$
acting on the left on $x$, we get $x'=gx$. Taking into account 
(\ref{eq:arisgroup}), they explicitly transform as:
 \begin{equation}
t'(\tau)=t(\tau)+b,\;   {\bi r}'(\tau)=R(\bbeta){\bi r}(\tau)+{\bi 
a},\end{equation}
as in (\ref{eq:Aristotle}) and also for the new degrees of freedom
 \begin{equation}
R(\balpha'(\tau))=R(\bbeta)R(\balpha(\tau)).
\label{eq:aristot}
 \end{equation}

The seven kinematical variables of our elementary particle
are now time $t$, position ${\bi r}$ and orientation $\balpha$. 
Our system can be interpreted as a 
point with a local Cartessian frame attached to it.
This local frame can rotate, and rotation of this frame is described by the evolution of the new
variables $\balpha$. 
Then the Lagrangian for this system
will be also a function of $\balpha$ and $\dot{\balpha}$, or equivalently 
of the angular velocity $\bomega$ of the moving frame. The homogeneity condition allows us to 
write $\widehat{L}$ as
 \begin{equation}
\widehat{L}=T\dot t+{\bi R}\cdot\dot{\bi r}+{\bi W}\cdot\bomega, 
 \end{equation}
where $T$ and ${\bi R}$, are defined as before (\ref{eq:Lagpoint}), 
and ${\bi W}=\partial\widehat{L}/\partial\bomega$. 

Now, total energy is $H=-T$, linear momentum ${\bi P}={\bi R}$, but the angular momentum takes the form
 \[ 
{\bi J}={\bi r}\times{\bi P}+{\bi W}.
 \] 
The particle, in addition to the angular momentum ${\bi r}\times{\bi P}$, has now a translation 
invariant angular momentum. The particle, a point and 
a rotating frame like the usual description of a rigid body, has spin ${\bi W}$. 

Nevertheless we have seen that while restricting ourselves to the Aristotle kinematical 
group we do not obtain generalized Lagrangians. All above Lagrangians depend only on the first order
derivative of variables ${\bi r}$ and $\balpha$.

The Principle of Inertia by Galileo
enlarges the Aristotle kinematical group $G_A$ to the whole Galilei group ${\cal G}$. 
The physical laws of dynamics must be independent of the 
uniform relative motion between inertial observers and this sets up a new 
kinematical group with a more complex structure. The action of the Galilei group is defined as
\[
t'(\tau)=t(\tau)+b,\quad   {\bi r}'(\tau)=R(\bbeta){\bi r}(\tau)+{\bi v}t(\tau)+{\bi a},
\]
which contains three new parameters ${\bi v}$, the relative velocity between observers.

In addition to the generators $H$, ${\bi P}$ and ${\bi J}$, the Galilei group has three new generators
${\bi K}$, which in the above group action are given by ${\bi K}=t\nabla$.

We see that once we have a larger group we can 
also enlarge, in an appropriate way, the kinematical variables of our 
point particle. The largest homogeneous space will contain as kinematical variables
the time $t$, position ${\bi r}$ and orientation $\balpha$, as the 
corresponding parameters of the Aristotle group but also the velocity of the particle ${\bi u}\equiv d{\bi r}/dt$
which comes from the corresponding group parameter ${\bi v}$.
Now the Lagrangian will be a function of these kinematical variables and their next order derivatives,
{\sl i.e.}, it must necessarily depend on the acceleration $d{\bi u}/dt$ of the particle. It is a Lagrangian 
that depends on the second derivative of the position variables ${\bi r}$. We thus get
a generalized Lagrangian for describing an elementary particle which will have a spin structure
that, in addition to the rotational spin as in the previous model, 
it has spin related to the zitterbewegung as we shall describe
in the next non-relativistic example.

\section*{A non-relativistic spinning particle}

We thus see that the most general non-relativistic particle will be described
by a Lagrangian which is a function of the variables $t,{\bi r},{\bi u},\balpha$ and 
$\dot{t},\dot{\bi r},\dot{\bi u},\dot{\balpha}$, being homogeneous of first degree
in terms of these last ones. It therefore can be written as
\begin{equation}
\widehat{L}=T\dot t+{\bi R}\cdot\dot{\bi r}+{\bi U}\cdot\dot{\bi u}+{\bi W}\cdot\bomega, 
\label{eq:gen}
\end{equation}
where we have replaced $\dot{\balpha}$ by the angular velocity $\bomega$ which is a linear function of it
and where here the new functions $U_i=\partial\widehat{L}/\partial\dot{u}_i$.

The Galilei group has non-trivial exponents\footnote{V. Bargmann, {\sl On unitary ray representations 
of continuous groups}, Ann. Math. {\bf 59}, 1 (1954). J.M. Levy-Leblond, {\sl Galilei Group and 
Galilean Invariance}, in E.M. Loebl, {\sl Group Theory and its applications}, 
Acad. Press, N.Y. (1971), vol. 2, p.~221.} and thus, according to (\ref{eq:alfa})
the Lagrangian is not invariant under the whole Galilei group but it transforms with the gauge function
\begin{equation}
\alpha(g;x)=\frac{m}{2}\left({\bi v}^2t+2{\bi v}\cdot R(\bbeta){\bi r}\right).
\label{eq:4.1.9}
\end{equation}
We see that if the group parameter ${\bi v}=0$ this gauge function vanishes so that
the non-invariance of the Lagrangian is coming only from its change under pure Galilei
transformations. 

Let us consider as a simpler example an elementary 
particle whose kinematical space is $X={\cal G}/SO(3)$. Any 
point $x\in X$ can be characterized by the seven variables 
$x\equiv(t,{\bi r},{\bi u})$, ${\bi u}=d{\bi r}/dt$, which are 
interpreted as time, position and velocity of the particle 
respectively. 

The Lagrangian will also depend 
on the next order derivatives, {\sl i.e.}, on the velocity which is already 
considered as a kinematical variable and
on the acceleration of the 
particle. Rotation and translation invariance implies that $\widehat{L}$ will be 
a function of only ${\bi u}^2$, $(d{\bi u}/dt)^2$ and ${\bi u}\cdot 
d{\bi u}/dt= d(u^2/2)/dt$, but this last term is a total time derivative 
and it will not be considered here. 

Let us assume that our elementary system is represented by the 
following Lagrangian, in terms of the kinematical 
variables and their derivatives, and in terms of some group invariant 
evolution parameter $\tau$,
 \begin{equation} 
\widehat{L}=\frac{m}{2}\frac{\dot{\bi r}^2}{\dot t}-
\frac{m}{2\omega^2}\frac{\dot{\bi u}^2 }{\dot t},
 \label{eq:n2}
 \end{equation} 
where the dot means $\tau$-derivative. 
Parameter $m$ is the mass of the particle because
the first term is gauge variant in terms of the gauge function (\ref{eq:4.1.9}) defined
by this constant $m$, while parameter $\omega$ of dimensions of 
time$^{-1}$ represents an internal frequency. It is the frequency of the 
internal zitterbewegung.   
If we consider that the 
evolution parameter is dimensionless, all terms in the Lagrangian have 
dimensions of action. Because the Lagrangian is a homogeneous function 
of first degree in terms of the derivatives of the kinematical 
variables, $\widehat{L}$ can also be written as 
 \begin{equation} 
\widehat{L}=T\dot t+{\bi R}\cdot\dot{\bi r}+{\bi U}\cdot\dot{\bi u},
\label{eq:n3}
 \end{equation} 
where the functions accompanying the derivatives of the kinematical 
variables are defined and explicitly given by 
 \begin{eqnarray} 
T&=&\frac{\partial\widehat{L}}{\partial\dot t}=-\frac{m}{2}\left(\frac{d{\bi r}}{ dt}\right)^2+
\frac{m}{2\omega^2}\left(\frac{d^2{\bi r} }{ dt^2}\right)^2,\nonumber\\ 
{\bi R}&=&\frac{\partial\widehat{L}}{\partial\dot{\bi r}}=m\frac{d{\bi r}}{ 
dt},\label{eq:n4}\\ {\bi U}&=&\frac{\partial\widehat{L}}{\partial\dot{\bi u}}=-
\frac{m}{\omega^2}\frac{d^2{\bi r}}{ dt^2}.\label{eq:n4bis}
 \end{eqnarray} 

In a time evolution description $\dot{t}=1$ $L$ it can be
written in terms of the three degrees of freedom
and their time derivatives as
 \begin{equation} 
L=\frac{m}{2}\left(\frac{d{\bi r}}{ dt}\right)^2-
\frac{m}{2\omega^2}\left(\frac{d^2{\bi r} }{ dt^2}\right)^2.
\label{eq:n1}
 \end{equation} 

Dynamical equations obtained from Lagrangian (\ref{eq:n1}) are: 
 \begin{equation} 
\frac{1}{\omega^2}\frac{d^4{\bi r}}{ dt^4}+\frac{d^2{\bi r}}{ dt^2}=0,
\label{eq:n5}
 \end{equation} 
whose general solution is: 
 \begin{equation} 
{\bi r}(t)={\bi A}+{\bi B}t+{\bi C}\cos\omega t+{\bi D}\sin\omega t,
 \label{eq:n6} 
 \end{equation} 
in terms of the 12 integration constants ${\bi A}$, 
${\bi B}$, ${\bi C}$ and ${\bi D}$. 

When applying Noether's theorem to the invariance of dynamical 
equations under the Galilei group, the corresponding constants of the 
motion can be written in terms of the above functions in the form: 
 \begin{eqnarray} 
\hbox{\rm Energy}\quad H&=&-T-{\bi u}\cdot\frac{d{\bi U}}{ dt},\label{eq:n71}\\ 
\hbox{\rm linear momentum}\quad {\bi P}&=&{\bi R}-\frac{d{\bi U}}{ dt}=
m{\bi u}-\frac{d{\bi U}}{ dt},\label{eq:n72}\\ 
\hbox{\rm kinematical momentum}\quad {\bi K}&=&m{\bi r}-{\bi P}t-{\bi U},\label{eq:n73}\\ 
\hbox{\rm angular momentum}\quad {\bi J}&=&{\bi r}\times{\bi P}+{\bi u}\times{\bi U}.
\label{eq:n7} 
 \end{eqnarray} 
It is the presence of the ${\bi U}$ function that distinguishes the 
features of this system with respect to the point particle case. We 
find that the total linear momentum is not lying along the direction of 
the velocity ${\bi u}$, and the spin structure is directly related to
the dependence of the Lagrangian on the acceleration.

We can think that the observable ${\bi Z}={\bi u}\times{\bi U}$ is the spin of the system.
We shall define the spin properly after the definition of the center of mass of the particle.
Nevertheless, magnitude ${\bi Z}$ looks like Dirac's spin operator, since taking the time derivative of
(\ref{eq:n7}), which is a constant of the motion, we get
\[
\frac{d{\bi Z}}{dt}={\bi P}\times{\bi u}
\]
similarly as the dynamical equation for the spin of a free particle obtained from Dirac's equation.

If we substitute the general solution (\ref{eq:n6}) in 
(\ref{eq:n71}-\ref{eq:n7}) we see in fact that 
the integration constants are related to the above conserved 
quantities 
\begin{eqnarray} 
H&=&\frac{m}{2}{\bi B}^2-\frac{m\omega^2}{2}({\bi C}^2+{\bi D}^2),\\ 
{\bi P}&=&m{\bi B},\\ 
{\bi K}&=&m{\bi A},\label{defK}\\ 
{\bi J}&=&{\bi A}\times m{\bi B}-m\omega{\bi C}\times{\bi D}.
 \end{eqnarray} 

We see that the kinematical momentum ${\bi K}$ in (\ref{eq:n73}) differs from the point particle 
case ${\bi K}=m{\bi r}-{\bi P}t$, in the term $-{\bi U}$, 
such that if we define the vector ${\bi k}={\bi U}/m$, with dimensions of length, 
then $\dot{\bi K}=0$ leads 
from (\ref{eq:n73}) to the equation: 
 \[
{\bi P}=m\frac{d({\bi r}-{\bi k})}{ dt},
 \] 
and ${\bi q}={\bi r}-{\bi k}$, defines the position of the center 
of mass of the particle. It is a different point than ${\bi r}$ and 
using (\ref{eq:n4bis}) is given by 
 \begin{equation}
{\bi q}={\bi r}-\frac{1}{m}{\bi U}={\bi
r}+\frac{1}{\omega^2}\;\frac{d^2{\bi r}}{ dt^2}.
\label{eq:n10}
 \end{equation} 
In terms of ${\bi q}$ the kinematical momentum takes the form
\[
{\bi K}=m{\bi q}-{\bi P}t.
\]

In terms of {\bi q} dynamical equations (\ref{eq:n5}) can be 
separated into the form: 
\begin{eqnarray} 
 \frac{d^2{\bi q}}{ 
dt^2}&=&0,\label{eq:n110}\\ \frac{d^2{\bi r}}{ dt^2}&+&\omega^2({\bi
r}-{\bi q})=0, 
\label{eq:n111} 
 \end{eqnarray} 
where (\ref{eq:n110}) is just eq. (\ref{eq:n5}) after twice
differentiating (\ref{eq:n10}), and Equation (\ref{eq:n111}) is (\ref{eq:n10})
after collecting all terms on the left hand side.

From (\ref{eq:n110}) we see that point ${\bi q}$ moves in a straight 
trajectory at constant velocity 
while the motion of point ${\bi r}$, given in (\ref{eq:n111}), is an isotropic harmonic motion
of angular frequency $\omega$ around point ${\bi q}$. 

The spin of the system ${\bi S}$ is now defined as the angular momentum
of the system but substracted the orbital angular momentum of its center of mass motion, {\sl i.e.},
 \begin{equation} 
{\bi S}={\bi J}-{\bi q}\times{\bi P}={\bi J}-\frac{1}{m}{\bi K}\times{\bi P}.
 \label{spin01}
 \end{equation} 
Since it is finally written in terms of constants of the motion 
it is clearly a constant of the motion, and its magnitude $S^2$ is also a Galilei invariant 
quantity that characterizes the system. In terms of the 
integration constants it is expressed as
 \begin{equation} 
{\bi S}=-m\omega\,{\bi C}\times{\bi D}.
 \end{equation} 
From its definition we get
 \begin{equation} 
{\bi S}={\bi u}\times{\bi U}+{\bi k}\times{\bi P}=-m({\bi r}-{\bi q})
\times\frac{d}{dt}\left({\bi r}-{\bi q}\right)=-{\bi k}\times m\frac{d{\bi k}}{dt},
 \label{eq:spin02}
 \end{equation} 
which appears as the (anti)orbital angular momentum of the relative 
motion of point ${\bi r}$ around the center of mass position ${\bi q}$, 
so that the total angular momentum can be written as
 \begin{equation} 
{\bi J}={\bi q}\times{\bi P}+{\bi S}={\bi L}+{\bi S}.
\label{angulJ}
 \end{equation} 
It is the sum of the orbital angular momentum ${\bi L}$ associated to the motion
of the center of mass and the spin part ${\bi S}$. For a free particle
both ${\bi L}$ and ${\bi S}$ are separately constants of the motion. We use the term (anti)orbital
to suggest that if vector ${\bi k}$ represents the position of a point mass $m$, the angular momentum
of this motion is in the opposite direction as the obtained spin observable. 
But as we shall see in a moment, vector ${\bi k}$ does not represent the position 
of the mass $m$ but rather the position of the charge $e$ of the particle.

Now one question arises: If ${\bi q}$ represents the center of mass position, then what 
position does point ${\bi r}$ represent? Point ${\bi r}$ represents 
the position of the charge of the particle. This can be seen by 
considering some interaction with an external field. The homogeneity 
condition of the Lagrangian in terms of the derivatives of the 
kinematical variables leads us to consider an interaction term of the 
form 
 \begin{equation} 
\widehat{L}_I=-e\phi(t,{\bi r})\dot t+e{\bi A}(t,{\bi r})\cdot\dot{\bi r},
 \label{eq:n13}
 \end{equation}
which is linear in the derivatives of the kinematical
variables $t$ and ${\bi r}$ and where the external potentials are only
functions of $t$ and ${\bi r}$. More general 
interaction terms of the form ${\bi N}(t,{\bi r},{\bi u})\cdot\dot{\bi 
u}$, and also more general terms in which functions $\phi$ and ${\bi A}$ also depend 
on ${\bi u}$ and $\dot{\bi 
u}$, can be considered. But this will be something different than an interaction with an 
external electromagnetic field. 

Dynamical equations obtained from $\widehat{L}+\widehat{L}_I$ are 
 \begin{equation} 
 \frac{1}{\omega^2}\frac{d^4{\bi r}}{ dt^4}+\frac{d^2{\bi 
r}}{ dt^2}=\frac{e}{ m} \left({\bi E}(t,{\bi r})+{\bi u}\times{\bi 
B}(t,{\bi r})\right),
 \label{eq:n14} 
 \end{equation}
where the electric field ${\bi E}$ and magnetic field ${\bi B}$ are 
expressed in terms of the potentials in the usual form, ${\bi E}=-
\nabla\phi-\partial{\bi A}/\partial t$, ${\bi B}=\nabla\times{\bi A}$. 
Because the interaction term does not modify the dependence of the 
Lagrangian on $\dot{\bi u}$, the function ${\bi U}=m{\bi k}$ has the same expression
as in the free particle case. Therefore the spin and the center of mass 
definitions, (\ref{eq:spin02}) and (\ref{eq:n10}) respectively, remain 
the same as in the previous case. Dynamical equations 
(\ref{eq:n14}) can again be separated into the form 
 \begin{eqnarray} 
 \frac{d^2{\bi q}}{ dt^2}&=&\frac{e}{ m}\left({\bi E}(t,{\bi r})+
{\bi u}\times{\bi B}(t,{\bi r})\right),\label{eq:n151}\\ 
\frac{d^2{\bi r}}{ dt^2}&+&\omega^2({\bi r}-{\bi q})=0,\label{eq:n152} 
 \end{eqnarray} 
where the center of mass ${\bi q}$ satisfies Newton's equations under 
the action of the total external Lorentz force, while point ${\bi r}$ 
still satisfies the isotropic harmonic motion of angular frequency 
$\omega$ around point ${\bi q}$. But the external force and the 
fields are defined at point ${\bi r}$ and not at point ${\bi q}$. It 
is the velocity ${\bi u}$ of point ${\bi r}$ that appears in the 
magnetic term of the Lorentz force. Point ${\bi r}$ clearly represents 
the position of the charge. In fact, this minimal coupling we have 
considered is the coupling of the electromagnetic potentials with the 
particle current, but the current $j_\mu$ is associated to the motion of 
a charge $e$ located at the point ${\bi r}$. 

This charge has an oscillatory motion of very high frequency $\omega$ 
that, in the case of the relativistic electron is $\omega=2mc^2/\hbar\simeq 
1.55\times10^{21}$s$^{-1}$. The average position of the charge is the 
center of mass, but it is this internal orbital motion, usually 
known as the {\bf zitterbewegung}, that gives rise to the spin structure and 
also to the magnetic properties of the particle.

When analyzed in the center of mass frame 
${\bi q}=0$, ${\bi r}={\bi k}$, the system reduces to a point charge whose 
motion is in general an ellipse, but if we choose $C=D$, and ${\bi 
C}\cdot{\bi D}=0$, it reduces to a circle of radius $a=C=D$, 
orthogonal to the spin. Then if the 
particle has charge $e$, it has a magnetic moment that according to the
usual classical definition is:~\footnote{\hspace{0.1cm}J.D. Jackson, {\sl Classical Electrodynamics}, 
John Wiley \& Sons, NY (1998), 3rd. ed. p.186.}
 \begin{equation} 
{\bmu}=\frac{1}{2}\int{\bi r}\times{\bi j}\, d^3{\bi r}=\frac{e}{2}\,{\bi k}\times\frac{d{\bi k}}{dt}= 
-\frac{e}{2m}{\bi S},
\label{eq:magneticmoment}
 \end{equation}  
where ${\bi j}=e\delta^3({\bi r}-{\bi k})d{\bi k}/dt$ is the current
associated to the motion of a point charge $e$ located at position ${\bi k}$.
The magnetic moment is orthogonal to the zitterbewegung plane
and opposite to the spin if $e>0$. It also has
a non-vanishing oscillating electric dipole ${\bi d}=e{\bi k}$,
orthogonal to ${\bmu}$ and therefore to ${\bi S}$ in the center of mass frame, 
such that its time average value 
vanishes for times larger than the natural period of this internal motion. 

Although this is a nonrelativistic example it is interesting to point out
and compare with Dirac's relativistic analysis of the electron,~\footnote{\hspace{0.1cm}P.A.M. Dirac, {\sl The Principles of Quantum 
mechanics}, Oxford Univ. Press, 4th ed. (1967).} in which
both, magnetic and electric momenta $\bmu$ and ${\bi d}$, respectively, appear 
giving rise to two possible interacting terms in Dirac's Hamiltonian. 

\section*{The gyromagnetic ratio}

The most general spinning particle under a Galilei Relativity Principle is the one with
a kinematical space $X={\cal G}$. As mentioned before, the most general Lagrangian
has the form (\ref{eq:gen}). What is important of its anlysis is the structure of the kinematical momentum
${\bi K}$ and angular momentum ${\bi J}$. They have the general form:
\[
{\bi K}=m{\bi r}-{\bi P}t-{\bi U},
\] 
as in the previous restricted example and where the observable ${\bi U}$, 
which is coming from the dependence of the Lagrangian on the acceleration,
is responsible of the separation between the center of mass and center of charge.
The zitterbewegung appears whenever we use generalized Lagrangians on the position variables
and the point ${\bi r}$ represents the center of charge of the particle. If ${\bi U}$ does not vanish
the particle has magnetic moment.

For the total angular momentum we get
\[
{\bi J}={\bi r}\times{\bi P}+{\bi u}\times{\bi U}+{\bi W}.
\]

We obtain again an angular momentum observable
\[
{\bi Z}={\bi u}\times{\bi U}+{\bi W},
\]
which satisfies, for the free particle, equation (\ref{eq:s}). This is the 
classical equivalent to Dirac's spin operator and contains two parts:
One ${\bi u}\times{\bi U}$ related to the zitterbewegung and therefore
to the magnetic moment of the particle, and another ${\bi W}$
related to the rotation of the particle as in the case of a rigid body, but which plays no role
in the dipole structure of the particle. 

A constant spin can be defined for the free particle if we substract from ${\bi J}$
the orbital angular momentum of the center of mass ${\bi q}\times{\bi P}$. In this case
the result is
\[
{\bi S}=-m{\bi k}\times\frac{d{\bi k}}{dt}+{\bi W},
\]
where as before ${\bi k}={\bi r}-{\bi q}={\bi U}/m$.

We see a clear kinematical feature: The magnetic moment is only related to the zitterbewegung
part of the spin. Therefore, from the experimental point of view, we can measure mechanical
and electromagnetic properties of the particle. When measuring the conserved spin of the particle, 
it is not possible to separate the measurement of both spin components. 
This implies that when expressed the magnetic moment in terms of the total 
spin their relationship is not the usual one and this produces 
the concept of gyromagnetic ratio $g$. The zitterbegung part 
of the spin only quantizes with integer values, because
it has the structure of an orbital angular momentum. Half 
integer values can come only from the rotation part of the spin.
This salient feature has recently been shown to lead to a gyromagnetic ratio $g=2$
for leptons and charged $W^\pm$ bosons whenever both components of 
spin contribute with their lowest admisible values.
\footnote{M. Rivas, J.M. Aguirregabiria and A. Hern\'andez, {\sl A pure kinematical explanation of 
the gyromagnetic ratio $g=2$ of leptons and charged bosons}, 
Phys. Lett. {\bf A 257}, 21 (1999).}
Deviations of $g-2$ are thus produced by radiation corrections as is shown 
in quantum electrodynamics.

\end{document}